\documentclass[twocolumn]{aastex631}
\usepackage{hyperref}
\hypersetup{
    colorlinks=true,
    linkcolor=blue,
    filecolor=magenta,      
    urlcolor=cyan,
}

\usepackage{xspace}
\usepackage{float}
\usepackage[version=4]{mhchem}

\usepackage[caption=false]{subfig}
\usepackage[T1]{fontenc}
\newcommand{\tess}{\textit{TESS} }
\newcommand{\kepler}{\textit{Kepler} }

\begin{document}
\nolinenumbers
\title{Assessing \textit{TESS's} Yield of Rocky Planets Around Nearby M Dwarfs}

\author[0000-0003-2404-2427]{Madison~T.\ Brady}
\affiliation{Department of Astronomy \& Astrophysics, University of Chicago, Chicago, IL 60637, USA}

\author[0000-0003-4733-6532]{Jacob L.\ Bean}
\affiliation{Department of Astronomy \& Astrophysics, University of Chicago, Chicago, IL 60637, USA}

\begin{abstract}
% Update: added lower radius limit in abstract.  TODO: check minimum radius in code, make sure it's consistent.  Make latex show bolding correctly.
Terrestrial planets are easier to detect around M dwarfs than other types of stars, making them promising for next-generation atmospheric characterization studies.  The \tess mission has greatly increased the number of known M dwarf planets that we can use to perform population studies, allowing us to explore how the rocky planet occurrence rate varies with host radius, following in the footsteps of past work with \kepler data.  In this paper, we use simulations to assess \textit{TESS's} yield of small ($0.5\,R_\oplus \,<\,R_p\,<\,2\,R_\oplus$) planet candidates around nearby ($d\,<\,30$\,pc) M dwarfs. We highlight the underappreciated fact that while \tess was indeed expected to find a large number of planets around M dwarfs overall, it was not expected to have a high planetary yield for the latest M dwarfs. Furthermore, we find that \tess has detected fewer planets around stars with $R_\star\,<\,0.3 R_\odot$ than even was expected (11 observed vs.\ $24\,\pm\,5$ expected).  We find evidence that the photometric noise of stars in the \tess bandpass increases with decreasing radius for M dwarfs. However, this trend cannot explain the observed distribution of planets. Our main conclusions are: (1) the planet occurrence rate likely doesn't increase, and may decrease for the latest M dwarfs; and (2) there are at least 17, and potentially three times that number, transiting planets around nearby late M dwarfs that still won't be detected by the end of \textit{TESS's} 4th year.

\end{abstract}


\section{Introduction}
\label{sec:intro}

M dwarfs provide an exciting opportunity to study terrestrial exoplanets.  Common exoplanet detection techniques (such as transit photometry and radial velocities) have signal amplitudes inversely related to the size of the host, making a planet easier to observe around an M dwarf than a larger, earlier-type star.  Additionally, M dwarfs have closer-in habitable zones \citep{Kopparapu13} than earlier-type stars.  This improves our chances of discovering habitable-zone planets around M dwarfs, given limited survey lengths.  This ``M Dwarf Opportunity'' is a key motivation of many ongoing and planned planet detection and characterization programs \citep{NAP25187}.

The \kepler mission \citep{Borucki10} found an increase in small planet occurrence as the stellar mass decreases \citep{Howard12, Dressing2015, Hardegree_Ullman_2019}.  However, given the small area of the sky surveyed, relatively few late-M dwarfs were observed.  Thus, these studies were primarily limited to quantifying planet occurrence around early- and mid-M dwarfs.  The K2 mission \citep{K2}, which re-purposed the \kepler spacecraft to observe targets along the ecliptic, has provided some constraints on planet occurrence around late-type stars.  However, studies using these data \citep[such as][]{Sagear20, Sestovic20} have only been able to place upper limits on rocky planet occurrence around ultracool stars.  Such stars are particularly important planet hosts due to the their significantly smaller size and closer-in habitable zones than even early type M dwarfs. To reinforce this point, we note that nine of the ten known systems of transiting planets around M dwarfs having $R_\star\,<\,0.35\,R_\odot$ and within 15\,pc of the solar system are the subject of atmospheric characterization observations with \textit{JWST} in Cycle 1 \citep[see the list in \S5 of][]{winters21}.

The \tess (\textit{Transiting Exoplanet Survey Satellite}) mission \citep{Ricker15} is an all-sky photometric survey that has been instrumental in the detection of new exoplanets that are ideal for follow-up observations.  With tens of thousands of M dwarfs in its target list \citep{Stassun18}, and several thousand planetary candidates already identified around stars of all types \citep{Guerrero21}, it vastly expands our ability to perform statistical studies of as-of-yet unprobed stellar and planetary populations.

\begin{figure}
    \centering
    \includegraphics[width=3.3in]{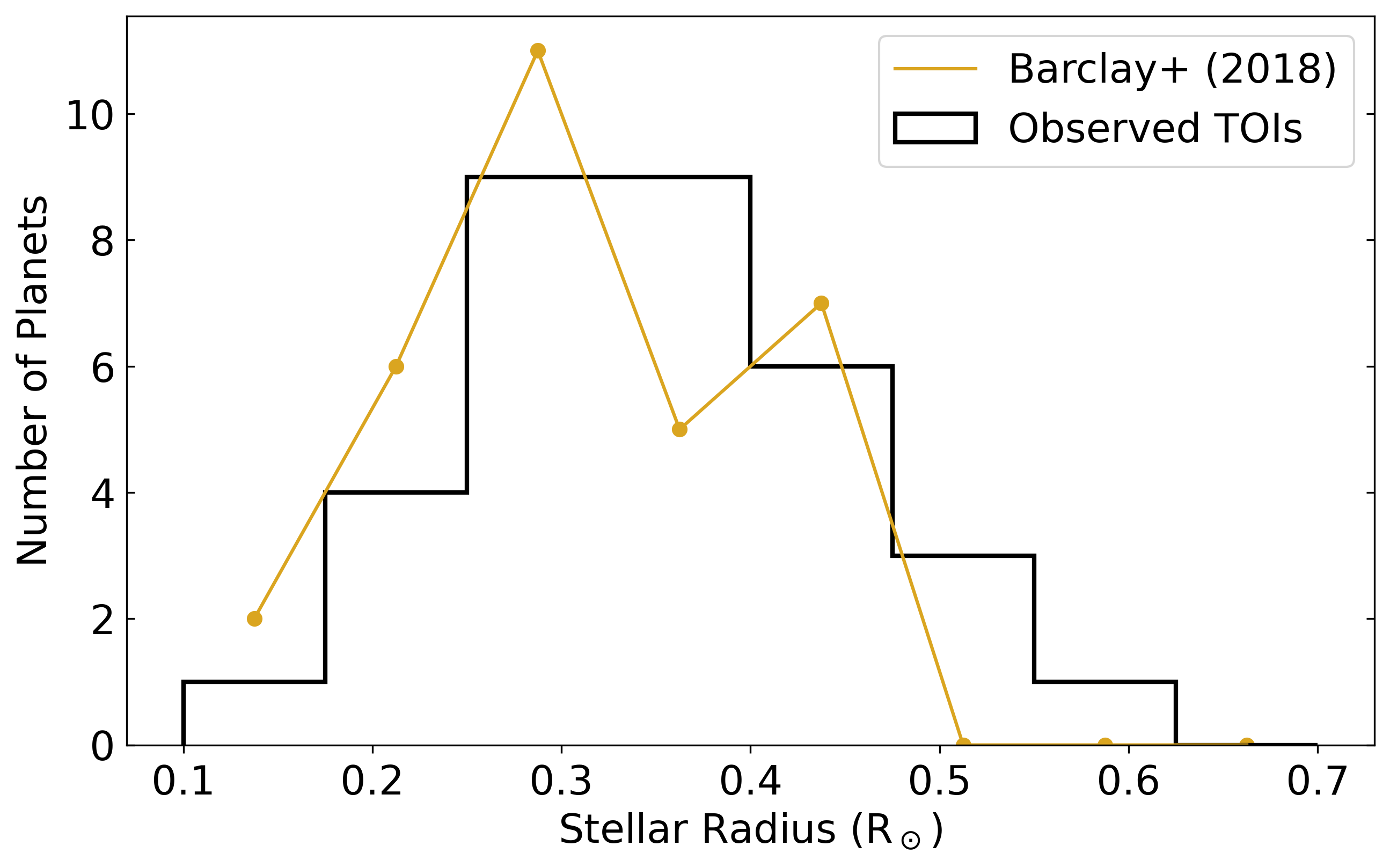}
    \caption{The distribution of rocky planets around M dwarfs as a function of stellar radius.  The histogram plots the observed planets from the \tess Prime Mission \cite{Guerrero21} in black, and the gold points are the results from the provided \cite{Barclay18} simulation.  The plotted sample only includes planets around M dwarfs within 30\,pc, $R_p\,<\,2 R_\oplus$, and $P\,<\,13.5$\,days (which corresponds to the length of half a single TESS sector).}
    \label{fig:barclay}
\end{figure}

The nearest ($d\,<\,30$\,pc) M dwarfs are brighter, making them more amenable to further study, and the \tess catalog likely includes a more complete sample of them than more distant populations.  Figure~\ref{fig:barclay} shows the distribution of small, short-period planet candidates (also known as TOIs, \tess Objects of Interest) around this subset of M dwarf hosts from the \tess prime mission \citep{Guerrero21}.  The distribution of planetary candidates with stellar radii has a clear peak around $0.3\,R_\odot$, with fewer planets detected both around smaller and larger stars.  The increase in detections with decreasing stellar radius likely reflects the increased detectability of these planets (the transit depth goes as $R_\star^{-2}$) and/or the large number of M dwarfs at low masses \citep{Winters15}.  The decrease in detections at radii smaller than $0.3\,R_\odot$ could reflect the decreased transit probability (which goes as $R_\star/a$) or the decreased brightness in the \tess bandpass of these dimmer and cooler stars.

It is necessary to include the known observational biases in any efforts to understand the distribution of planet candidate properties and to determine whether or not additional phenomena, such as stellar activity and planet occurrence, are influencing our observations.  Stellar activity, which can meaningfully increase observed stellar photometric noise, appears to be more prevalent in later-type M dwarfs than in warmer, more massive M dwarfs \citep{West15}.  Meanwhile, there is observational evidence that the occurrence rate of rocky planets is higher around smaller M dwarf stars \citep{Hardegree_Ullman_2019, Sabotta21}.

Before the launch of the \tess mission, \cite{Barclay18} ran a suite of simulations to estimate the potential yield of the \tess mission in Sectors 1 -- 26, using a noise model purely based upon the \tess magnitude of the star.  They didn't account for the potential effects of stellar activity and they assumed a single overall occurrence rate for M dwarfs, so their simulations provide us with an appropriate baseline of the planets we should expect to observe given what we currently understand about the biases described above.  They provided the results of a single run of these simulations as a list of stars and their detected planets.  Figure~\ref{fig:barclay} compares the observed rocky, short-period M dwarf planets from the prime mission to a comparable sample (with the same cuts) from the \cite{Barclay18} simulation.

Overall, if we assume Poisson $\sqrt{n}$ errors, the simulation results do not seem to disagree at a greater than one sigma level within most bins.  The simulations also seem to broadly reproduce the shape of the observed distribution of planet candidates, with a peak in detections at intermediate radii.  It also predicts a similar number of planets in the overall sample (33 candidates vs.\ 31 in the simulated results).  However, a closer examination shows that the simulation appears to systematically underestimate the detection of planets around the largest of stars ($R_\star\,>\,0.5\, R_\odot$) and overestimate detections around the smallest of stars ($R_\star\,<\,0.3\,R_\odot$).  While this only represents a single instance of a suite of simulations, it points towards the possibility that we are observing fewer planets around very low-mass stars than what was anticipated before the \tess launch.  However, the \cite{Barclay18} results do not give us the statistical strength necessary to draw any concrete conclusions. 

This motivates us to perform a new suite of simulations, focusing specifically on nearby M dwarfs, using the actual \tess yield of planet candidates and the list of stars that were observed. \tess has surveyed many more stars since the end of Year 2, which should lend our simulations additional statistical strength.  In addition, we run a large number of instances of these simulations, to give us a sense of how our results vary.  We choose to index our models using the stellar radius, as the stellar radius is a fundamental parameter of interest in transit observations, and it directly affects the observed planetary signal.  We also include models of radius-dependent (as radius closely traces mass around main-sequence M dwarfs) stellar activity and planet occurrence, to see if these meaningfully affect the distribution of observed planets.

If the observed discrepancy in rocky planet detections at $R_\star\,<\,0.3\,R_\odot$ is due to a decrease in planet occurrence or an increase in stellar activity, it could have implications with regards to future searches for planets around these tiny stars.  A decrease in planet occurrence at small stellar radii could also provide us with insight into how planets form around M dwarfs.  \cite{Mulders_2021} theorized that the rocky planet occurrence distribution may peak at intermediate M dwarf masses, using a pebble drift and accretion model.  They found that the increase in giant planet formation around larger stars tended to inhibit the formation of rocky planets.  Meanwhile, tiny host stars simply lacked the necessary material to form large numbers of planets.  \cite{Pan21} also found a dependency between stellar mass and planet occurrence, resulting in a decrease in planet occurrence around the smallest of stars ($M_*\,<\,0.2\, M_\odot$) in most formation scenarios.

In this paper, we use our simulations to assess the yield of rocky planet candidates around nearby M dwarfs, especially those around stars with $R_\star\,<\,0.3\,R_\odot$. To do this, we select a volume-limited sample of \tess M dwarfs, simulate planets around them, and compare the observed planet distribution with that found by \textit{TESS}.

In \S\ref{sec:methods}, we describe the \tess sample and how we simulated observing them, while \S\ref{sec:results} discusses our simulation results and how they compare to the observed planet distribution.  We conclude in \S\ref{sec:conclusions}.

\section{Methods}
\label{sec:methods}

\subsection{\tess Sample Selection}
\label{ssec:tess_sample}

M dwarfs are dim, and rocky planets produce shallow transit signals.  Thus, for the reasons described in \S\ref{sec:intro}, we limit our sample to stars from the \tess Input Catalog (TIC) that are within 30 parsecs.  We also only include stars with $T_{\mathrm{eff}}\,<\,4,000$\,K and $log\,g\,>\,4.0$, to avoid sample contamination from early-type stars, pre-main-sequence stars, and giants.  A sample consisting primarily of main sequence dwarfs will have a tight correlation between mass and radius, allowing us to describe our models in terms of stellar radius, which is more directly related to the planetary transit properties than the stellar mass.  Additionally, pre-main-sequence stars tend to have larger radii and be more active than main sequence stars \citep[see, e.g.,][]{Scholz07}, which would complicate our activity simulations if they were included in the sample.

We took our stellar parameters from version 8.2 of the TIC, which is described in \cite{Paegert21} and represents the most up-to-date version of the list. This version incorporates updates to version 8 of the TIC \citep[described in][]{Stassun19} for the purpose of identifying non-existent objects.

Version 8 of the TIC incorporates Gaia DR2 \citep{GaiaDR2} parallaxes and colors.  The stellar temperatures are derived using an empirical $T_{\mathrm{eff}}$-\textit{Gaia} color relation.  Radii are found using the Stefan-Boltzmann relation, with reddening and bolometric corrections.  Masses are inferred using a $T_{\mathrm{eff}}$-mass relation for unevolved stars, as long as the star has a radius consistent with not being in the giant branch.  The stellar $log\,g$ is calculated using the determined mass and radius.  Given the likely inclusion of subgiants in the sample, the masses and $log\,g$ values of any potentially evolved stars determined this way may be inaccurate.  However, we primarily describe the stars in terms of radius in this study, and any potential mass error on these targets are only impactful indirectly through our $log\,g\,>\,4.0$ cutoff.

Additionally, nearly all (97\%) of the M dwarfs in this version of the TIC that made our parameter cuts are included in the Second Cool Dwarf Catalog (CDC2), which was an update to the Cool Dwarf List described in \cite{Muirhead18}.  This catalog was built from a curated list of cool objects with color, brightness, and error cutoffs described in \cite{Stassun19}.  For these targets, effective temperature, radius, and mass were derived using magnitude and color relationships from \cite{mann13}, \cite{Mann15}, and \cite{Mann19} respectively.  For the objects in CDC2, the stellar parameters derived in this way were included in the TIC instead of those calculated with the default procedures.

After incorporating distance, temperature, and surface gravity cutoffs, our sample contains 4,910 stars from the TIC, with radii ranging from $R_{\star}$\,=\,0.1 -- 0.9\,$R_\odot$.  Around half of these stars have $R_{\star}\,<\,0.3\,R_\odot$.  Of this list, 3,761 are included in the Sectors 1 -- 43 observed two-minute cadence target list on \href{https://tess.mit.edu/observations/target-lists/}{the \tess website}.  The remaining stars either have longer cadences or are yet to be observed.  We found that omitting the small sample of long-cadence targets had a negligible effect on our results, and thus include them.  We also include the stars that have not been observed yet in our sample for the sake of forecasting future results.

To compare with our simulations, we made use of the most recent list of TOIs through Sector 43 (from the TOI release portal as of 2021 November 30), with similar cuts performed on the host stars ($T_{\mathrm{eff}}\,<\,4,000$\,K, $log\,g\,>\,4.0$, $d\,<30$\,pc). We also limited our sample to planet candidates with periods less than half of a single \tess sector (due to its 27-day long observing sectors,
\textit{TESS}'s results are incomplete for planets with longer orbital periods) and radii between $0.5\, R_\oplus$ and $2\,R_\oplus$ (to limit our scope to rocky planets with acceptable detection statistics).  Our final sample includes 41 planet candidates, with host radii $R_\star \,=\,0.15$ -- $0.6\,R_\odot$.

% Update- clarifying TOIs vs planets 
We assume that our TOI sample is 100\% complete for the detection threshold of the \tess pipeline (7.1$\sigma$, see next subsection). We also assume that there are no false positives in this limited sample. We note that two-thirds of the 33 planet candidates in our sample that come from the Prime Mission (and thus have had sufficient time for further study) have already been statistically validated and/or confirmed via radial velocity measurements in the literature. Furthermore, none of the candidates in our sample have been identified as false positives to date. Therefore, we refer to the planet candidates as planets below for brevity. 

\subsection{Modeling Planet Observations}
\label{ssec:pl_model}

After selecting our sample of stars, we performed a suite of simulations in order to model the number of planets we would expect to detect around them, given what we know from \cite{Dressing2015} about how these planets are distributed and the sensitivity of \textit{TESS}.  To model this, we first generated a population of planets with random radii and orbital parameters around the stars from \S\ref{ssec:tess_sample}.  After this, we determined geometrically which stars hosted transiting planets.  Finally, we determined the number of observed planet transits and the transit signal-to-noise ratio (SNR) based on the host star's characteristics and on-sky location.  Planets with a high SNR and multiple transits observed were marked as ``detectable'' by \textit{TESS}.  A more detailed description of this process follows.

\subsubsection{Modeling The Planets}
\label{sssec:pl_par}

At the beginning of each simulation, we assigned each host star in the \S\ref{ssec:tess_sample} sample a number of planets according to a Poisson distribution with mean $\lambda$:

\begin{equation}
    N_{pl} = \mathrm{Pois}(\lambda)
\end{equation}

In the nominal model, also used by \cite{Barclay18}, we allowed $\lambda\,=\,2.5$, which is representative of the expected number of planets around M dwarfs with $R_{p}$\,=\,1 -- 4\,$R_\oplus$ and $P\,<\,200$ days according to the analysis by \cite{Dressing2015}, using \kepler data.  However, we also wanted to explore the possibility that planet occurrence is not constant across the entire range of main sequence M dwarf radii, and thus also created additional models in which $\lambda$ is a function of $R_\star$, $\lambda(R_\star)$.  To explore the possibility of a radius-dependent planet occurrence, we considered two different functional forms of $\lambda(R_\star)$:

\begin{itemize}
    \item $\lambda\,=\,aR_\star^{p}+b$ (power-law model).  This is an effort to model the increase in 
    planet occurrence with decreased M dwarf mass observed in \cite{Hardegree_Ullman_2019} and \cite{Sabotta21}, though the exact functional form of this relationship is currently unknown.
    \item $\lambda \,=\, A e^{-(R_\star-\mu)^2/(2 \sigma^2)} + b$ (Gaussian).  This model reproduces the increase in planet occurrence at intermediate stellar mass, and decrease at very low stellar mass predicted by \cite{Mulders_2021}.  It also matches up with the observed decrease in planet occurrence around late M dwarfs, observed by K2 \citep{Sagear20, Sestovic20}.
\end{itemize}

We allow the exact values of the parameters in these functions to vary according to the data, and fit them by-eye in our results, as we are more interested in getting an idea of whether or not these functions show promise with regard to explaining the data than we are in determining the exact values of these parameters.  A significantly expanded TOI list (to reduce the effect of the Poisson errors) and a full completeness analysis would be likely be necessary to properly determine the planet occurrence rate as a function of radii with some degree of statistical rigor, which is beyond the scope of this work. 

After determining the number of planets per star, we calculated the physical and orbital parameters (radius, period, mid-transit time, inclination, eccentricity, and angle of periastron - $R_p$, $P$, $t_0$, $i$, $e$, and $\omega$ respectively) of each planet by sampling from random distributions that are described in the following paragraphs.

The planet radii and periods are drawn from the 2D distributions taken from \cite{Dressing2015}, with radii between 0.5 -- 4.0\,$R_\oplus$ and $P\,<\,200$\,days.  \cite{Dressing2015} provide their occurrence data by offering the expected number of planets per star as a function of both planetary radius and period, using stellar properties from \cite{Huber14}.  As we are interested in determining the probability that an individual planet has a given radius and period, we normalize the whole table by dividing each radius/period bin by the sum of all of the individual occurrence rates in the entire table.  This creates a table in which the value in each entry represents the probability that a given planet has a radius and period within the defined bin limits. We then determine the radius/period bin each planet falls in according to a random draw from this 2D distribution.  Then, we draw the precise radius and period of each planet assuming that the radius and period vary uniformly between the bin limits.  We note that this is only an approximation of the actual R vs P probability distribution. As we are primarily interested in looking at rocky planets using \tess data (which is incomplete at long periods), we only consider rocky (0.5 -- 2.0$\,R_\oplus$), short-period ($P\,<\,27/2$\,d) planets in our final visualizations.

The mid-transit time ($t_0$) for each planet was drawn from a random uniform distribution with limits based on the period.  The cosine of the line-of-sight inclination is drawn from a random uniform distribution between zero and one, with the assumption that, in the case that a star has multiple planets, they are all coplanar.  We note that this method of simulating inclinations doesn't take into account the known Kepler dichotomy, in which roughly half of planetary systems are expected to have single planets or multiple with high mutual inclinations, and the other half are expected to possess multiple planets that are nearly coplanar \citep{Ballard16}.  We found that fully relaxing the coplanarity assumption (such that every planetary inclination is fully independent) only affected our predictions within each stellar radius bin by less than $0.1\,\sigma$, so we do not expect this exclusion to meaningfully impact our results.

We allow the planetary eccentricity to vary as the distribution given in \cite{Xie16}, who found that single-transiting planets are fit by a Rayleigh distribution with a significantly higher mean than multi-transiting planets ($\approx\,0.32$ vs.\ $\approx \,0.04$).  The argument of periastron is also drawn from a random uniform distribution between $\omega\,=\,0$ -- $2 \pi$.  We found that fixing $e$ and $\omega$ at zero in these simulations only affected the overall number of planets detected (and not the shape of the distribution), and even then only caused a change on the order of 20\% or less in each radius bin.  As we lack sufficient evidence to conclude that these planets would have been circularized, we do not fix the eccentricity.

Once the planet parameters are determined, we can calculate the specific observable characteristics of their transits.  This is elaborated upon in the following section.

\subsubsection{Calculating the Transit Parameters}

After simulating a sample of planets, we need to determine which of these planets would actually transit their host stars.  The impact parameter $b$ can be calculated using the formula described in, for example, \cite{Winn2010}:

\begin{equation}
    b = \frac{a~\mathrm{cos}i}{R_\star} \bigg(\frac{1-e^2}{1+e~\mathrm{sin}\omega}\bigg)
\end{equation}

where $a$ is the planet's semi-major axis, which can be calculated given the planet's orbital period and the (known) mass of the host star.  If $b\,>\, (1 + R_P/R_\star)$, the planet does not transit.  Given that we are looking at a sample of rocky planets (and thus $R_p/R_\star$ tends to be small), we consider planets with $b\,<\,1$ as having observable transits in our simulations for the sake of simplicity.  Given the planet parameters described in \S\ref{sssec:pl_par}, we find that we underestimate the total number of planetary transits by only around 5\% by ignoring these strongly-grazing transits.  However, highly grazing transits tend to have shallower transit depths and thus tend to be more difficult to observe in general, so this simplification will probably have an even weaker impact on our final results.

If a planet is transiting, it is straightforward to calculate the transit depth $\delta\,=\,(R_p/R_\star)^2$, as the star's radius is known and the planet's radius is simulated.  Due to \textit{TESS}'s large pixel size, light from nearby stars can dilute the transit light curve from the host, making the fractional transit depth (and thus the derived planet radius) much smaller.  To correct for this, similar to \cite{Barclay18}, we also include the effect of dilution due to nearby stars in our depth determination, multiplying the observed depth by a factor of $1/(1+r_{\mathrm{cont}})$, where $r_{\mathrm{cont}}$ is the flux contamination ratio from the TIC.  

We added in the effects of limb-darkening, as \cite{Heller2019} found that neglecting the effects of limb-darkening could cause depth underestimates on the order of 30\% for M dwarf planets in \textit{Kepler} data.  For each star in our sample, we find their limb-darkening coefficients using the tables of values from \cite{Claret17}, assigning each star an $a$ and $b$ in accordance with what was calculated for the star that was the most similar (in terms of $T_{\mathrm{eff}}$ and $log\,g$) in the sample.  We used the coefficients that were calculated assuming a quasi-spherical model and computed with the least square method, and limited ourselves to solar-metallicity models with a microturbulent velocity of $V\,=\,2$ km/s, as those were what was available for the full range of low-temperature models.  We used the PHOENIX-DRIFT model coefficients for stars with $T_{\mathrm{eff}}\,<\,3050$ K and the PHOENIX-COND model coefficients for hotter stars.

With these coefficients in hand, the transit depths were calculated directly adapting some of the geometric formulae from \cite{Heller2019}, assuming a quadratic limb-darkening law.  Multiplying that by the contamination factor, we get the final formula for $\delta$:

\begin{multline}
    \delta = \bigg(\frac{R_p}{R_\star}\bigg)^2 \\ 
    \times \frac{1-a_{\mathrm{LD}}\big(1-\sqrt{1-b^2}\big)-b_{\mathrm{LD}}\big(1-\sqrt{1-b^2}\big)^2}{1-\frac{a_{\mathrm{LD}}}{3}-\frac{b_{\mathrm{LD}}}{6}} \\
    \times \frac{1}{1+r_{\mathrm{cont}}}
\end{multline}

where $a_{\mathrm{LD}}$ and $b_{\mathrm{LD}}$ are the quadratic limb-darkening coefficients.  Limb darkening has a weak impact on our results, typically resulting in slight (sub-5\%) increases in overall detection counts when compared to simulations without limb darkening.

The transit duration, $T_{\mathrm{dur}}$, is the amount of time that the planet spends in-transit.  Longer transits are typically easier to detect, especially on targets with longer cadences.  To calculate the transit duration, we use the formula discussed in \cite{Winn2010}:

\begin{equation}
    T_{\mathrm{dur}} = \frac{P}{\pi} \mathrm{sin}^{-1}\bigg( \frac{\sqrt{(R_\star+R_P)^2 - (b~R_\star)^2}}{a}\bigg).
\end{equation}

For all planets considered to be transiting, we can calculate $\delta$ and $T_{\mathrm{dur}}$.  The following section describes how these parameters, along with the characteristics of the star, can be used to determine whether or not the transit would be be observed and classified as a signal of interest by \textit{TESS}.

\subsubsection{Determining Observability}
\label{sssec:observability}

We can only claim a planet detection if at least two transits of the planet are observed during the \tess run.  If we only observe a single transit, it is impossible to fully constrain a planet's period, and it is difficult to rule out the possibility that the flux dip is caused by some other effect (such as a starspot or another star).  

For each star in the \tess sample, we determined which sectors it was observed in using the \texttt{TESS-point} software, provided by \cite{Burke20}.  \texttt{TESS-point} compares the known location of the star with the known region of the sky surveyed in each sector, and allows for us to determine both the number of sectors each star was observed for as well as the actual specific sectors each star was observed in.  The total number of observed transits ($n_T$) were determined by finding, based on $t_0$ and $P$, how many transits would be observed given the known sectors the host was observed.  Given the known start and stop times of each sector, such a calculation is straightforward to perform for the sake of comparing the host stars to the observed planets for Sectors 1 -- 43.

However, as of the time of writing, the \tess website only provides accurate start and stop times of sectors up through the first orbit of Sector 47 (each sector consists of two orbits), so we must estimate the future observation times if we are interested in performing forecasts regarding future planet detections.  To do so, we made use of the expected sector midpoint times included in the \texttt{tess-point} software up through Sector 55 and determined the start and stop times of each future sector orbit by assuming that each sector lasts roughly 12.5 days (the average length of a TESS orbit in sectors 1-47) and that there is a data gap of roughly 1.2 days between each orbit (once again using the average values from previous sectors).  We stress that we only used these estimations in the occasion that actual observational data was not available, which is only relevant for our future forecasts.

The second factor that determines the observability of a transit is its SNR, where higher-signal transits are more likely to be observable.  The SNR itself is related to $\delta$, $T_{\mathrm{dur}}$, and the photometric noise.  

To describe this noise, we used the one-hour Combined Differential Photometric Precision noise ($\sigma_{\mathrm{CDPP}}$, calculated purely from the host star T magnitude (listed in the TIC) using \texttt{ticgen} \citep{Stassun18}. We note that the code's outputs seem to be consistent with the lower envelope of the \tess magnitude vs.\ CDPP noise found on-sky in the \href{https://tasoc.dk/docs/release_notes/tess_sector_01_drn01_v02.pdf}{Sector 1 \tess Data Release Notes}, indicating that 
\texttt{ticgen} gives us an estimate of \textit{TESS}'s minimum noise (which is equivalent to its read noise plus its photon noise).  However, it appears that many targets have an observed $\sigma_{\mathrm{CDPP}}$ much higher than this estimate.  It is possible that these deviations from the expected noise could be related to the physical characteristics of the stars themselves.  Any additional source of noise in a star's light curve can make its planets more difficult to detect, influencing the observed planet yield.  If certain types of stars tend to be noisier, this can result in systematic deviations between simulations and observations in plots like Figure~\ref{fig:barclay}.

Stellar activity can contribute meaningfully to the noise in a photometric light curve \citep[see, e.g.,][]{Carpano03, Oshagh18}.  However, not all stars are equally active: M dwarfs spin down more slowly \citep{Delfosse98} and have more fractional X-ray flux at given rotational periods \citep{Kiraga07} than other types of stars.  If smaller stars tend to be more active, we may detect fewer planets around smaller stars due to their reduced SNR compared to what the magnitude-derived noise would suggest.  If the scatter in the relationship between observed $\sigma_{\mathrm{CDPP}}$ and the star's T magnitude is correlated with the star's radius (where smaller stars result in higher noise), activity could explain the fact that we're observing fewer planets than expected around the smallest of host stars.  As the CDPP noise model doesn't account for this factor, we include a slight modification to the $\sigma_{\mathrm{CDPP}}$ that we use to calculate the transit SNR, adding an additional radius-dependent ``activity noise'' term, $\sigma_{A}$. This takes the form

\begin{equation}
\label{eqn:fr}
    \sigma_{\mathrm{CDPP, obs}}(T, R_\star) = \sqrt{\sigma_{\mathrm{CDPP, shot}}(T)^2 + \sigma_A(R_\star)^2}.
\end{equation}

To estimate the magnitude and functional form of $\sigma_A$, we downloaded (via \texttt{lightkurve}) the \tess PDCSAP light curves of every star in our sample for which they were available.  Overall, 3,672 light curves were available for download- slightly fewer than the number of targets \tess has observed so far (3,860, estimated using \texttt{tess-point}). This difference in number is small enough that it is unlikely to meaningfully impact our results.  We then calculated the targets' $\sigma_{\mathrm{CDPP}}$ directly from the light curves, using \texttt{lightkurve}, which performs an estimate using the ``sgCDPP proxy algorithm'' described by \cite{Gilliland11} and \cite{VanCleve16}.  Low frequency-signals were first removed from the light curve using Savitsky-Golay filtering, in which a second order polynomial was fit centered on each point in the time series with a window length of two days.  After 5$\sigma$ clipping, the CDPP noise was estimated as the standard deviation of a running mean over a one hour window in the smoothed data.  After using this method to determine the observed CDPP noise in the light curve, we used Equation~\ref{eqn:fr}, letting $\sigma_{\mathrm{CDPP,short}}=\sigma_{\mathrm{CDPP,ticgen}}$ and $\sigma_{\mathrm{CDPP, obs}}=\sigma_{\mathrm{CDPP,lightkurve}}$, solving for $\sigma_A$. 

\begin{figure*}
    \centering
    \includegraphics[width=6in]{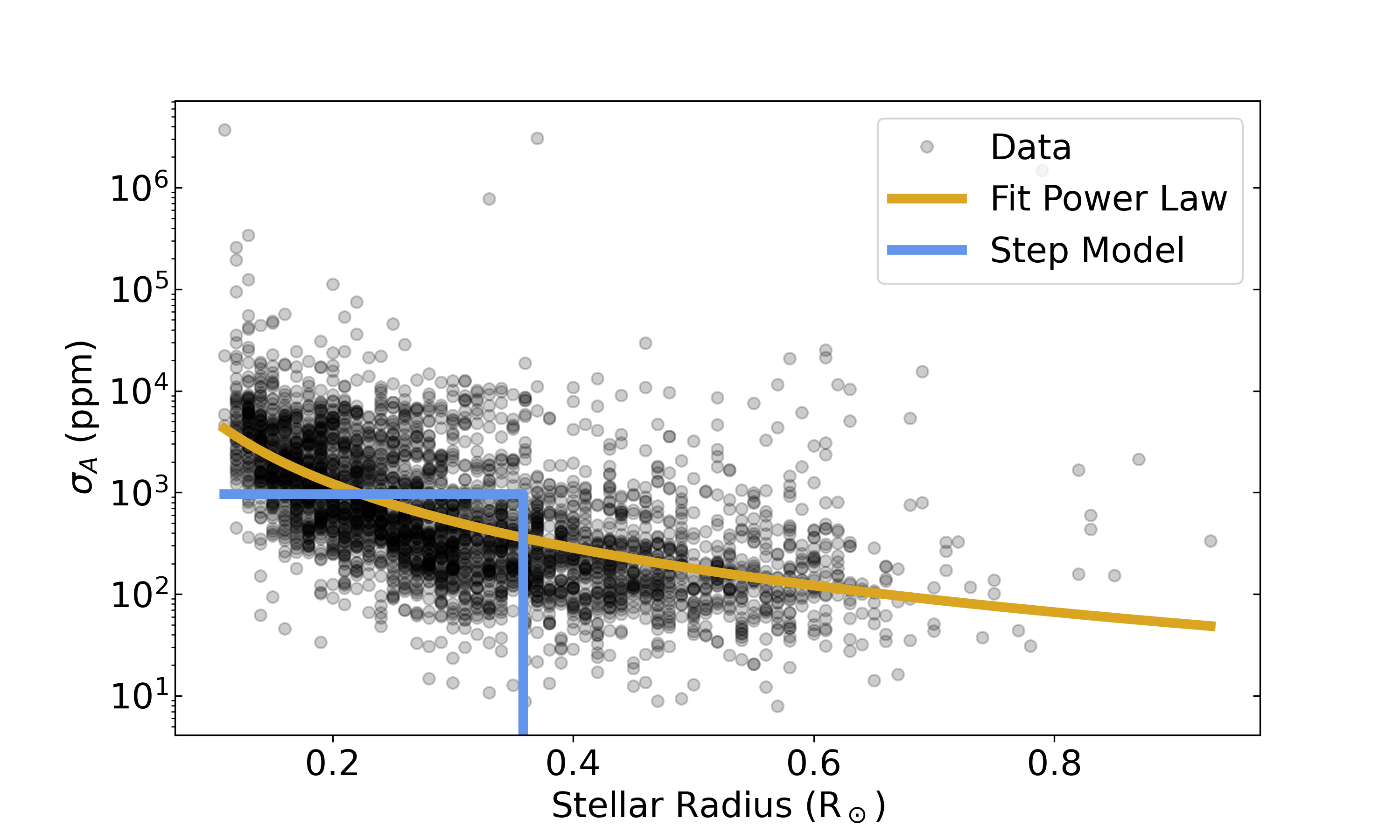}
    \caption{Additional noise, $\sigma_A$, as a function of stellar radius.  $\sigma_A$ is calculated by comparing the 1-hour CDPP noise determined directly from light curves using \texttt{lightkurve} to the noise calculated with \texttt{ticgen}.  The best-fit power-law model is included in gold.  A step model with a cutoff at $0.36~R_\odot$ (corresponding to where we expect M dwarfs to transition into a fully convective regime) is included in blue.}
    \label{fig:cdpp}
\end{figure*}

Figure~\ref{fig:cdpp} shows our resulting values for $\sigma_A$, as a function of stellar radius.  We determined a best-fit power-law model for the relationship by transforming the data into log-log space and fitting a linear model, finding that the relationship is best explained by

\begin{equation}
\label{eqn:fit_f}
    \mathrm{log} \bigg( \frac{\sigma_A}{\mathrm{ppm}} \bigg) = \big(-2.09 \pm 0.05\big)\times \mathrm{log} \bigg( \frac{R_\star}{R_\odot} \bigg) + \big(3.73 \pm 0.06\big).
\end{equation}

This trend supports our theory that the photon and read noise are insufficient to understand the \tess data, and that it is necessary to include an additional term to estimate planet yield.  The stellar radius does have a direct impact on the observed stellar magnitude through the size of the emitting surface, however, so we caution against using this relationship to describe stellar activity without a more careful analysis of the relationships between stellar radius, temperature, brightness, and $\sigma_{\mathrm{CDPP}}$.  However, that being said, we can investigate how a radius-dependent noise model impacts the observed planet distribution by using this data, alongside theory, to motivate possible functional forms of $\sigma_A$ for our simulations.  We choose to examine three forms:

\begin{itemize}
    \item $\sigma_A\,= 0$ (uniform model).  This model coincides with the photon noise plus the read noise of \textit{TESS}, and assumes no radius dependence in the $\sigma_{\mathrm{CDPP}}$.
    \item $\sigma_A\,=\,a$ when $R_\star\,<$ cutoff, $\sigma_A\,= \,b$ when $R_\star\,>$ cutoff (step model).  This model assumes that there is some cutoff mass below which the $\sigma_{\mathrm{CDPP}}$ of the stars increases.  This emulates the divide between fully convective M dwarfs and other types of stars, which are believed to have a different type of dynamo mechanism. \cite{Newton16} places the location of this cutoff around $0.35\,M_\odot$ (corresponding to $0.36\,R_\odot$ in our sample).  This aligns somewhat with the observed dramatic increase in $\sigma_R$ at small stellar radii, but it doesn't capture the precise functional form.  While this doesn't match the observed data as well as power-law model shown in Figure~\ref{fig:cdpp}, it will be interesting to explore whether or not this much simpler model can do a decent job at explaining the data.  In our simulations, we will model this convective cutoff by letting $\sigma_A\,=\,0$ for stars with radii above $0.36~R_\odot$, and $\sigma_A\,\approx \,970$ at smaller radii, corresponding to the mean of the logarithms of the calculated $\sigma_A$ in this regime.
    \item $\sigma_A\,=\,aR_\star^{p}$ (power-law model).  This model, with $p<0$, is meant to reproduce the increasing observed fraction of active M dwarfs with decreasing radius that is observed in \cite{West15}.  As active stars tend to be noisier than quiet stars, we would expect to see an increase in $\sigma_{\mathrm{CDPP}}$ around smaller stars.  This matches up with the roughly power-law dependence in $\sigma_A$ observed in Figure~\ref{fig:cdpp}.  We can input this relation directly into our simulations using Equation~\ref{eqn:fit_f}.
\end{itemize}

Once we have the CDPP noise, we can estimate the SNR of a given transit with 

\begin{equation}
\label{eqn:snr}
    \mathrm{SNR} = \frac{\delta}{\sigma_{\mathrm{CDPP}}} = \bigg(\frac{R_p}{R_\star}\bigg)^2 \frac{1}{\sigma_{\mathrm{CDPP, 1hr}}}\sqrt{n_T T_{\mathrm{dur}}}
\end{equation}

where $T_{\mathrm{dur}}$ is in hours, for proper comparison to the one-hour CDPP noise.  We consider planets with SNR $>$ 7.1 and $n_T\,>\,2$ as having been detected, to mimic the detection threshold of the \tess pipeline (see the \href{https://tasoc.dk/docs/release_notes/tess_sector_01_drn01_v02.pdf}{\tess Data Release Notes}).  Adopting more conservative detection thresholds of SNR $>$ 10 and $n_T\,>\,3$ causes a roughly 20-30\% decrease in overall observations but does not affect the shape of our resulting occurrence curves.  

After determining which planets would be detectable, we repeat the simulations, starting from \S\ref{sssec:pl_par}.  For each noise and planet occurrence model, we repeat the simulations 10,000 times (increasing the number of simulations beyond this point did not seem to affect our results). The planet detection statistics are generated by finding the mean and standard deviation of the number of planets ``observed'' during each run.

\section{Results}
\label{sec:results}
% A lot of this section has been rewritten.
In this section, we describe the results of our suite of simulations.  Our nominal simulation reproduces the rise and then fall of the number of detected planets in the M dwarf regime as a function of stellar radius. This reinforces the expectation from the \citet{Barclay18} results that \tess would find fewer planets around the latest M dwarfs than mid M dwarfs.  However, they overpredict the total number of rocky M dwarf planets, especially those around the smallest stars.  Below, we describe these results in more detail and explore the possibility that stellar activity or a nonuniform planet occurrence could explain these deviations.  We then use these results to try explain why \tess is missing planets, and provide estimates as to the number of rocky M dwarf planets that could be found by the end of \tess Year 4.

% Old version:  Our results are shown in Figures~\ref{fig:noise} and \ref{fig:occ}.  The uncertainty in our simulations appear to be primarily dominated by Poisson $\sqrt{N}$ noise in each bin.  Our nominal simulation reproduces the rise and then fall of the number of detected planets in the M dwarf regime as a function of stellar radius. This reinforces the expectation from the \citet{Barclay18} results that \tess would find fewer planets around the latest M dwarfs than mid M dwarfs.  However, it overpredicts the total number of rocky M dwarf planets, especially those around the smallest stars.  We describe this model and its results in more detail, as well as possible explanations for our overprediction, below.  We also use our simulation to break down the reasons for \tess missing transiting planets around late M dwarfs.

\subsection{A Stellar Radius-Dependent Noise Model}
% Cuts made- placed explanation of stellar activity in methods
\label{ssec:noise}

\begin{figure*}
    \centering
    \includegraphics[width=6in]{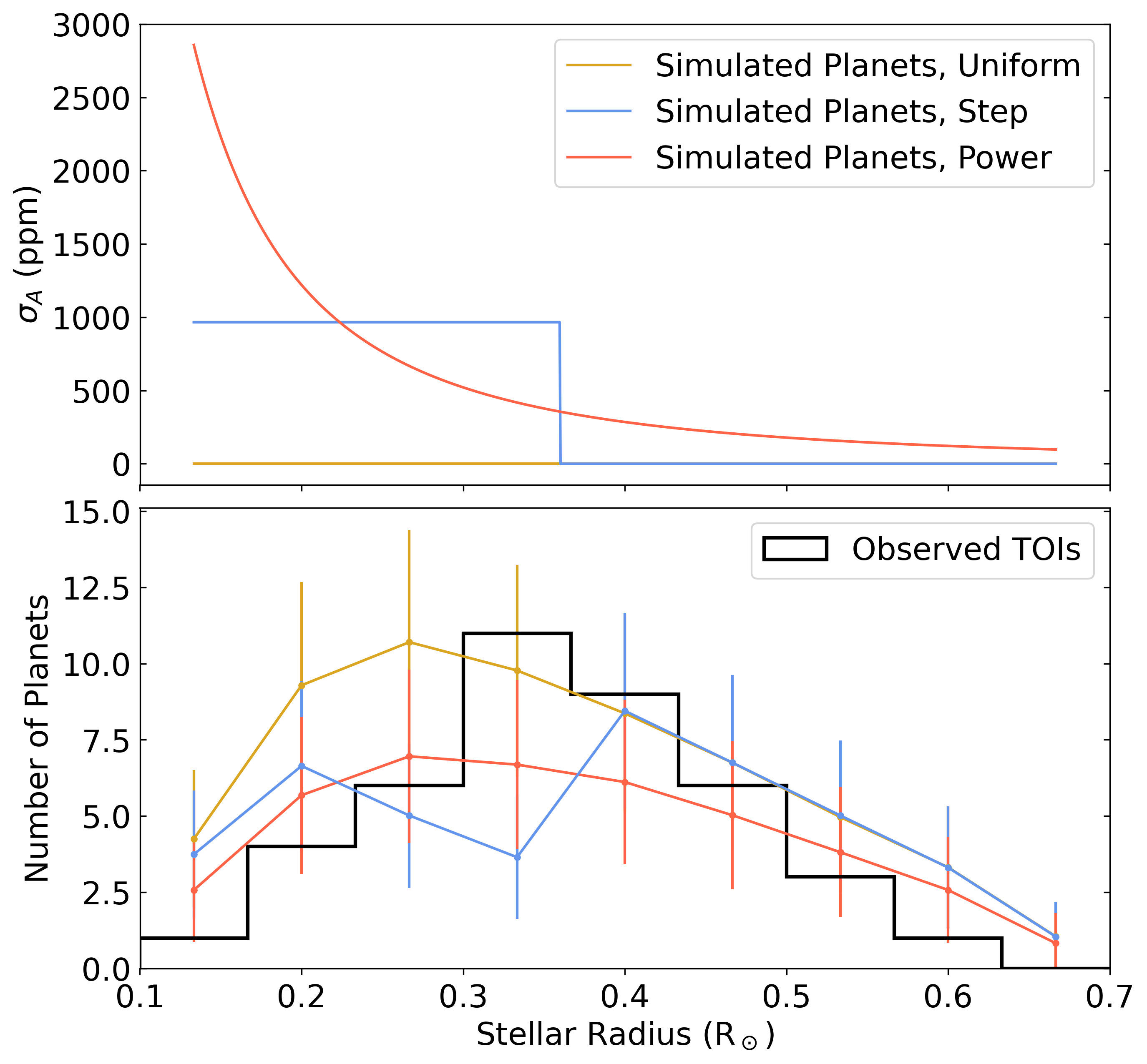}
    \caption{\textbf{Top:} The three forms (uniform, step, and power-law) of the additional noise term, $\sigma_A$, as a function of stellar radius. \textbf{Bottom:} The number of simulated planetary detections (with errors) around nearby M dwarfs with the given noise models, colored to reflect the models shown in the top panel of the figure.  The step and power-law noise models are informed by fits to $\sigma_A$, described in \S\ref{sssec:observability}.  A histogram of the observed \tess planets from sectors 1-43 is included for comparison.}
    \label{fig:noise}
\end{figure*}

As described in \S\ref{sssec:observability}, it is more difficult to detect planets around more active stars, which could easily result in a drop in observed planets.  As later M dwarfs are more likely to be active \citep[see][]{West15}, this could likely manifest as a decrease in detected planets around very small hosts that cannot be explained through other means.  To represent this increased difficulty in planet detection, we have included a radius-dependent term $\sigma_A$, meant to represent activity, into the determination of the light curve SNR.  The functional forms of $\sigma_A$ fed into our simulation are shown in the top panel of Figure~\ref{fig:noise}.  Following the methodology in \S\ref{ssec:pl_model}, we generate 10,000 samples of planet observations for each of these three noise models and plot their distribution in the bottom panel of Figure~\ref{fig:noise}.  In this section, we let the mean planet occurrence $\lambda\,=\,2.5$ planets per star, reflecting the results from \cite{Dressing2015} for planets around M dwarfs.  

The uniform model with $\sigma_A\,=\,0$ represents our prior assumptions about the noise model of this planet distribution.  It agrees with the number of observed planets within $1-2\sigma$ within each individual bin, but overpredicts the overall number of planets around M dwarfs (41 observed vs.\ 58\,$\pm$\,9 predicted).  The simulated planet distribution also has a different shape from the observations, appearing flatter and peaking at a lower radius ($\approx\,0.25 R_\odot$ in simulations versus the observed $\approx\,0.35 R_\odot$).  The model  confidently overpredicts planets around stars with $R_\star\,<\,0.3\,R_\odot$ (11 observed vs.\ $24\,\pm\,5$ predicted) and slightly overpredicts those around stars with $R_\star\,>\,0.5\, R_\odot$ (4 observed vs.\ $9\,\pm\, 3$ predicted).  This could indicate a degree of pipeline incompleteness around the smallest stars or just an overall underestimation of noise around M dwarfs, though the latter statement does not explain why our simulations \textit{under}predict the number of planets around mid-M stars.

The step model, with a cutoff at $0.36\,R_\odot$ (which corresponds to the convective cutoff at $M_\star \, \approx \, 0.35\, M_\odot$ motivated by \cite{Newton16}), takes on the value of $\sigma_A\,=\,0$ at large radii, assuming the \texttt{ticgen}-derived noise is accurate in this regime.  We adopt the $\sigma_A \approx 970$ for $R\, < \, 0.36\,R_\odot$ fit from the \texttt{lightkurve} data in \S\ref{sssec:observability}, and find that the step model allows for a slightly closer estimation of the number of planets around the smallest stars (15\,$\pm$\,4 planets predicted vs.\ the known 11).  However, it cannot explain the overpredictions around larger stars and produces a double-peaked distribution that doesn't match the \texttt{lightkurve} $\sigma_{\mathrm{CDPP}}$ measurements. However, given the significant $\sqrt{N}$ errors on each point, we would need to observe a much larger number of planets to definitively rule out such a model.

The power-law model, produced using the parameters from Equation~\ref{eqn:fit_f}, has slightly more accurate predictions than the nominal model, estimating 16\,$\pm$\,4 planets around the $R_\star\,<\,0.3\,R_\odot$ stars and 8\,$\pm$\,3 planets around the $R_\star\,>\,0.5\,R_\odot$ stars (compared to observed counts of 11 and 4, respectively).  However, this model is incapable of reproducing the distinctive observed planet peak around $0.35\,R_\odot$, resulting in a much broader and flatter distribution that underestimates the number of planets in the intermediate regime and overestimates the number of planets in the tails.  We note that directly inputting the calculated value for $\sigma_A$ from \texttt{lightkurve} for each object for which it is known produces a similar result to this simpler power-law model- a flatter, broader distribution with a peak at a lower radius than the observations, though this noise inclusion also still resulted in an overestimation of the number of planets at low radii.  

The fact that the power-law model underestimates the peak while improving the fit to the tails implies that one way to accurately reproduce the ``sharpness'' of the observed peak is with some sort of V- or U-shaped noise model with a minimum around $0.3\,R_\odot$.  We note that M dwarfs have been found to have increasing $H_\alpha$ and $R'_{HK}$ indices with radius above $0.3\,R_\odot$  \citep{Robertson13, Astudillo-Defru17}, so this could indicate a cutoff between a radiative regime and a fully-convective regime.  However, this peaked noise model doesn't align with our observations in Figure~\ref{fig:cdpp}, motivating a search for an alternative explanation for the observed planet distribution.

\subsection{A Non-Uniform Occurrence Around Small Stars}
% Cuts made- placed explanation of occurrence in methods
\label{ssec:occ}

\begin{figure*}
    \centering
    \includegraphics[width=6in]{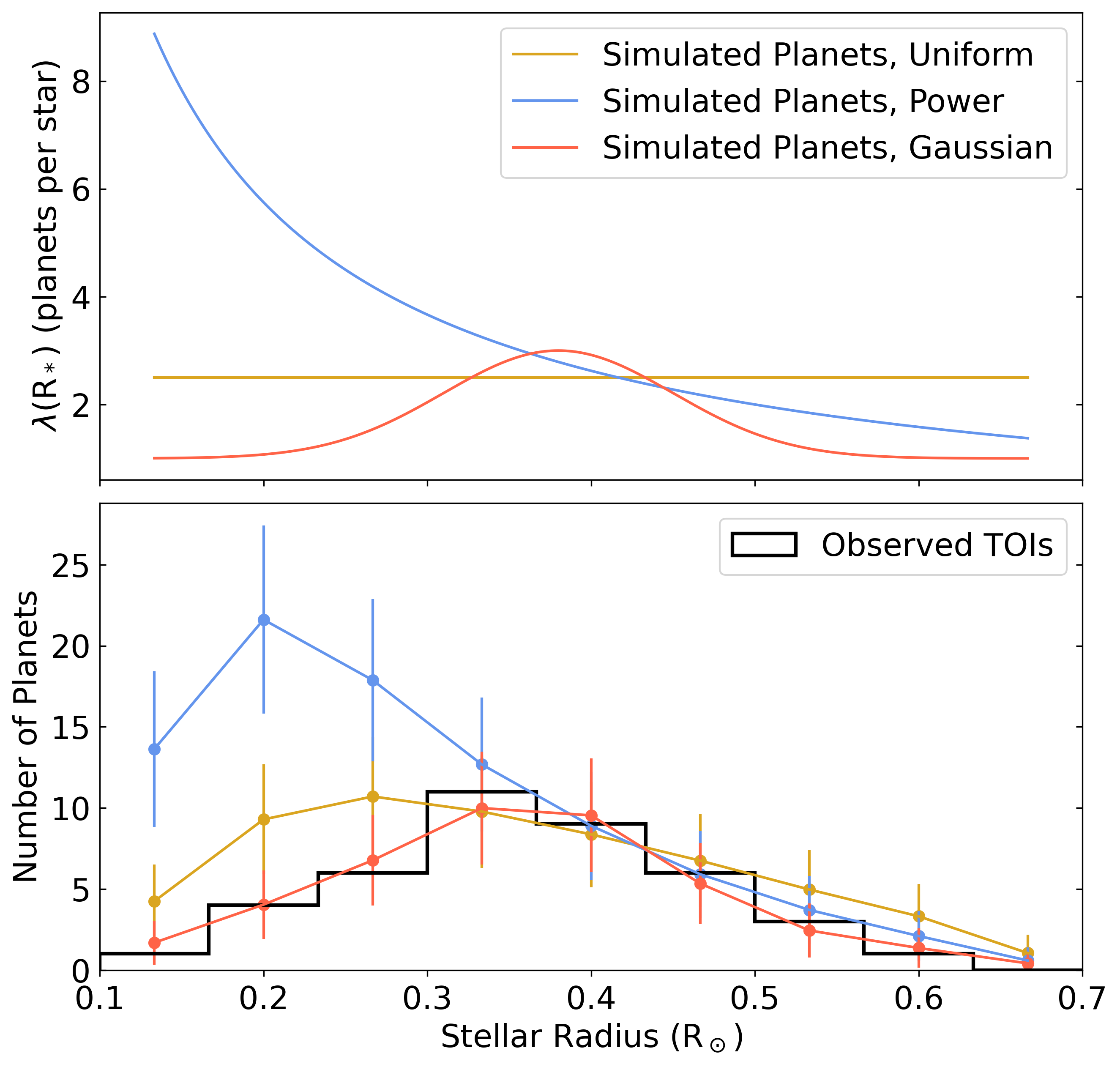}
    \caption{\textbf{Top:} The three forms (uniform, power-law, Gaussian) of the occurrence model, $\lambda$, as a function of stellar radius when $\sigma_A\,= \,0$.  \textbf{Bottom:} The number of simulated planetary detections (with errors) around nearby M dwarfs with the three different planet occurrence models, colored to reflect the models shown in the top portion of the figure.  The power-law model was tuned by-eye to match the observed number of planets around large M dwarfs, and the gaussian model was tuned to roughly match the center, width, and height of the observed distribution. A histogram of the observed planets from sectors 1-43 is included for comparison.}
    \label{fig:occ}
\end{figure*}

As we were unable to fully explain the observed deviations between our nominal simulation and the \tess observations with activity-dependent noise models, it is possible that the actual occurrence rate of planets around M dwarfs has some dependency on their radii.  As described in \S\ref{sssec:pl_par}, we can explore this effect by generating the number of planets per star using a Poisson distribution with a $R_\star$-dependent mean.  The various functional forms of $\lambda$, as well as our simulation results, are shown in Figure~\ref{fig:occ}.  When testing these models, we fixed $\sigma_A\,=\,0$.  We also checked to see what happened when allowing $\sigma_A$ to be equivalent to the values calculated directly by \texttt{lightkurve} for the stars for which such data was available, and otherwise equal to the value of $\sigma_A$ derived by Equation~\ref{eqn:fit_f}.  We found fixing $\sigma_A\,=\,0$ did not change the conclusions of this section, and only affected the particular derived parameters for each fit.  We primarily quote results from the $\sigma_A\,=\,0$ models, but will include discussions on how models with the \texttt{lightkurve} $\sigma_A$ change the fit parameters.

Each parameter in the functional forms of $\lambda(R_\star)$ were fit by eye.  In the future it may be possible to perform a more statistically rigorous fit, but the large errors on our measurements (as well as the lack of a full completeness analysis) makes such a task beyond the current scope of this paper. We are primarily interested in determining which functional forms of planetary occurrence rate match up with the current data from \textit{TESS}.  A description of the nominal simulation, with a constant planet occurrence rate and $\sigma_A\,=\,0$, is included in \S\ref{ssec:noise}.

Parameters from a power-law fit of the occurrence rate around M3, M4, and M5 stars from \cite{Hardegree_Ullman_2019} and $\sigma_A\,=\,0$ produced a very poor fit to the observed planet distribution, significantly underestimating the detection of planets around $R\, > \,0.3\,R_\odot$ stars (30 observed vs.\ 10\,$\pm$\,3 predicted planets) and overestimating those around $R\, < \,0.3\,R_\odot$ stars (11 observed vs.\ 34\,$\pm$\,7 predicted planets).  Using the $\sigma_A$ calculated in \S\ref{sssec:observability} instead (when available) further reduces the number of planets observed around larger stars, widening the divide even further.  This is possibly due to the small number of data points with large error bars included in the \cite{Hardegree_Ullman_2019} fit, as well as the fact that their sample includes a different subset of planetary periods and radii than our selected planet distribution.  

As these methods tended to significantly underestimate the observed number of planets, especially around the largest stars, we adopted $\sigma_A=0$ and estimated by-eye the power-law parameters that could accurately reproduce the number of planets observed around stars with $R\,>\, 0.3\, R_\odot$.  We found that a model with $a\,\approx\,1.25$, $p\,\approx\,-1$, and $b\,\approx\,-0.5$ can predict the observed planet rate for $R_\star\,>\,0.3\, R_\odot$ accurately, but dramatically overpredicts the number of planets around stars with $R\, < \,0.3 R_\odot$ (11 observed vs.\ 53\,$\pm$\,9 predicted planets), and peaks at even smaller radii.  Using the \texttt{lightkurve} estimate of $\sigma_A$ results in a similarly poor fit to the observations, requiring a steeper slope in planetary occurrence ($p\approx-1.5$ as opposed to $p=-1$) to offset the increase in noise from the $\sigma_A$ term and fit the observed planet rate around larger stars.  However, this model similarly overpredicts the number of planets around the smallest stars.  Even though it appears like the increase in noise around smaller stars would reduce the overestimation of planets at small radii, it appears that it is impossible to produce reasonable estimates of the number of planets around both smaller and larger M dwarfs with a power-law occurrence model, either overestimating the number of planets around the smallest stars or underestimating the number around larger ones.  Below $R\, \approx \, 0.3 \, R_\odot$, the occurrence rate likely flattens or even decreases towards smaller stars.

The Gaussian model can be tuned to neatly reproduce the exact shape and amplitude of the observed planet distribution, providing support for the theory that there is a decrease in planet occurrence around the smallest stars.  We find that this model can reproduce the overall number of planets around M dwarfs (41 observed vs.\ 42 $\pm$ 7 predicted), as well as the number of planets around $R_\star \, < \, 0.3 \, R_\odot$ stars (11 observed vs.\ 12 $\pm$ 4 predicted) and those around $R_\star \, > \, 0.5 \, R_\odot$ (4 observed vs.\ 4 $\pm$ 2 predicted) within less than one sigma.  While the exact functional form of the planet occurrence with stellar radius is unknown (the large error bars on the individual bins make a precise determination difficult), a model that includes an occurrence peak at $\approx0.38\,R_\oplus$ at 3 planets per star can explain our observations.  As stars with radii $\sim\,0.35\,R_\odot$ are common and relatively bright, this could explain why we observe an average $\lambda\,=\,2.5$ planets per star \citep{Dressing2015} in transit data even if many M dwarfs have a substantially lower planet occurrence rate. 

This finding also agrees roughly with the results from \cite{Sestovic20}, who found an upper limit on the $P\,=\,$1 -- 20 d rocky occurrence rate of 1.14 planets per star, indicating a decrease in occurrence at low host masses.  \cite{Sagear20} also found a decrease in occurrence around very small host stars, though both of these surveys had poor sensitivity with regards to $R_p\,<\,2\,R_\oplus$ planets.

Our Gaussian model also seems to agree (within a factor of a few) with the inward and reversed migration models for terrestrial planetary formation from \cite{Pan21}, which found similar planet occurrence rates around M dwarfs.  Their reversed migration model found a peak at intermediate radii, and subsequent decrease in occurrence, though their calculated occurrence with stellar radius relation is flatter than what we observe and peaks around $0.2\,M_\odot$, which corresponds to radii significantly below the observed peak in the planet distribution.

Adopting the values of $\sigma_A$ from \texttt{lightkurve} and our power-law fit allows for a similarly good fit with a Gaussian that peaks at a similar (or slightly higher) stellar radius ($\approx0.42\,R_\oplus$) and an amplitude at its peak around 6 planets per star as opposed to 3 planets per star.  This does not necessarily contradict the estimation of $\lambda \, \approx \, 2.5$ planets per star from \cite{Dressing2015} for similar reasons as the Gaussian model with $\sigma_A\,=\,0$, but does contradict, e.g., \cite{Pan21}, which, with theoretical formation and migration models, did not find such a large-amplitude peak in planetary occurrence at intermediate radii, or indeed such a high planetary occurrence around any systems unless all of the planets were formed in-situ.

In general, the Gaussian model fit provides evidence in support of the model in \cite{Mulders_2021}, which showed that the low pebble flux around 0.1 -- 0.2 $M_\odot$ stars results in fewer planets.  However, our observed peak in planetary formation rate is around $0.4\,R_\odot$, which is slightly lower than their estimate of a peak around $0.5\,M_\odot$.  This could potentially be explained with their relatively coarse grid of models or the general challenges associated with making predictions for planet formation. 

These results do not necessarily disagree with those in \cite{Sabotta21}, which found an increase in planet occurrence below $0.34\,M_\odot$, as their analyzed subsample included more high-mass M dwarfs than low-mass ones.  This could weight the observed planet occurrence rate in $M_*\,<\,0.34\,M_\odot$ objects towards that observed around stars with $M_*\approx0.34\,M_\odot$, which likely fall around the peak in occurrence, while planets around $M_*\,>\,0.34\,M_\odot$ stars have more representation from the low-occurrence tails.  

Some combination of a complex noise and occurrence model (such as an increase in both noise and planet occurrence at low radii) could be responsible for the shape of the planet distribution, though this doesn't seem to be supported by our findings using the empirical noise estimates in \S\ref{ssec:noise}.  We found that using the more accurate values of $\sigma_A$ from \texttt{lightkurve} (or merely the power-law fit to $\sigma_A$) when available didn't affect any of our conclusions from this section, only the precise best-fitting parameters of the individual models for $\lambda$.

\subsection{Number of Remaining Unobserved Planets}
\label{ssec:forecast}

As our simulations (especially the Gaussian model) seem to be reasonably capable of reproducing the observed distribution of planets with M dwarf radius, we can use them to perform estimates of the number of planets that we are currently missing.  In \S\ref{sssec:observability}, we describe the precise cutoffs that determine whether or not a planet is observable (SNR $>$ 7.1 and $n_T\,\geq\,2$).  However, instead of merely counting up the number of observable planets, we can also keep track of the planets that transited but weren't observed for enough transits, be it due to their low SNR, long periods, or merely because their host has yet to be observed by \textit{TESS}.  Using the forecast from \texttt{TESS-point}, we can also estimate the number of planets we can expect to observe by the end of Sector 55 (September 2022). 

We can thus split the planets into three mutually exclusive categories:

\begin{itemize}
    \item The star has been observed by \tess and the planet does transit its host, but only zero or one transits have been observed.  This usually happens for long-period planets.
    \item The transit has been observed, but the SNR is too low.  This is more likely to affect planets with small radii, or very noisy stars.
     \item The host star is in a region of the sky that has not been observed with \tess yet.
\end{itemize}

We also consider a fourth category that is not necessarily mutually exclusive with the other three- the number of planets that isn't observable currently, but would be by the end of Year 4, based on the current sector plan.  This gives us insight into the number of planets that we can expect to be added to the sample of nearby rocky M dwarf planets in the future.  

For the sake of determining these counts, we use both the nominal model and the fit Gaussian model.  The nominal model appears to provide reasonable (within 1 -- 2 $\sigma$) estimates of the observed planet distribution, but seems to overestimate the number of planets around the smallest and largest stars in our sample.  Meanwhile the Gaussian occurrence model provides a very good estimate in our sample but is somewhat poorly constrained. We also use the models assuming $\sigma_A\,=\,0$ models, as reproducing the \tess observations using the \texttt{lightkurve} $\sigma_A$ requires the planetary occurrence rate around $0.3\,R_\odot$ stars to be extremely high.  We calculate the number of unobserved planets with each model individually- the actual number likely falls near or between these two model predictions. 

To explore how \tess performs on long-period planets, we broaden our limits in this case to include all planets with $P\,<\,200$ days.  The number of planets in each of the previously described categories for the nominal and Gaussian model are shown in Table~\ref{tab:missing_planets}. 

\begin{table*}
%\begin{tabular}{ccccc}
%\caption{Simulated Numbers of Planets Missed by \textit{TESS}}
\centering
\begin{tabular}{>{\centering}p{0.12\textwidth}>{\centering}p{0.12\textwidth}>{\centering}p{0.12\textwidth}>{\centering}p{0.12\textwidth}>{\centering\arraybackslash}p{0.12\textwidth}}
\hline
\textbf{R$_*$} & \textbf{$n_T<2$} & \textbf{SNR$<7.1$} & \textbf{Host Unobserved} & \textbf{Observable by S55}  \\ 
\hline
\multicolumn{5}{c}{\textbf{Nominal Model}} \\
\hline
0.10-0.30 R$_\odot$ & 11.0 $\pm$ 4.1 & 23.5 $\pm$ 5.6 & 15.5 $\pm$ 5.3 & 5.3 $\pm$ 2.6 \\
0.30-0.50 R$_\odot$ & 10.7 $\pm$ 4.1 & 21.4 $\pm$ 5.2 & 15.5 $\pm$ 5.3 & 5.5 $\pm$ 2.6 \\
0.50-0.70 R$_\odot$ & 4.4 $\pm$ 2.6 & 8.7 $\pm$ 3.4 & 6.2 $\pm$ 3.4 & 2.0 $\pm$ 1.6 \\
\hline
\multicolumn{5}{c}{\textbf{Gaussian Model}} \\
\hline
0.10-0.30 R$_\odot$ & 5.6 $\pm$ 2.7 & 11.8 $\pm$ 3.7 & 7.8 $\pm$ 3.4 & 2.7 $\pm$ 1.7 \\
0.30-0.50 R$_\odot$ & 10.8 $\pm$ 4.2 & 21.4 $\pm$ 5.3 & 15.7 $\pm$ 5.3 & 5.5 $\pm$ 2.6 \\
0.50-0.70 R$_\odot$ & 2.0 $\pm$ 1.6 & 3.9 $\pm$ 2.1 & 2.8 $\pm$ 2.0 & 0.9 $\pm$ 1.0 \\

\end{tabular}
\caption{The total number of simulated transiting rocky $R_p\,<\,2R_\oplus$ planets around $d\,<\,30$ pc M dwarfs that would be missed by \textit{TESS}, given our detection limits of $n_T \,\geq \,2$ and SNR $>$ 7.1.  The results are broken down based on why each individual planet is not observable by the end of Sector 43: either the planet doesn't transit twice during the observed sectors, the transit SNR is too low, or the host star has not been observed by \tess yet. The final column includes the number of planets that were not observable by our metrics by the end of Sector 43, but may be observable given data through the end of Sector 55.  The top table includes estimates from the nominal model, where $\sigma_A\,=\,0$ and $\lambda\,=\,2.5$, while the bottom table includes estimates based on the tuned Gaussian occurrence rate from Figure~\ref{fig:occ}.}
\label{tab:missing_planets}
\end{table*}

From this table, we can see that around 18-26 planets transit their host stars at a high SNR, but too few transits are observed to make a planet detection.  Such planets likely have periods long enough that they can be missed by a 27 day \tess sector.  Unfortunately, it will be hard to revisit such planets with \textit{TESS}, as \tess follows a fairly strict schedule (up through the end of Year 4) in terms of which parts of the sky it observes when.  In the short term, we cannot request to have \tess return to a part of the sky where a star was observed with a single transit to try to find a second transit, though follow-up is possible in the future.  Such systems could also have ground-based follow-up to search for a second transit, but it is very difficult to constrain the orbital period of a planet based on a single transit, as even with a very accurate transit duration measurement there are degeneracies between the impact parameter, limb darkening, orbital period, eccentricity, and $\omega$.

Meanwhile, 37-54 planets have had multiple transits observed via \tess, but at low SNR, due to a dim host star or their own small radii.  There is a possibility that, with more sophisticated light curve processing or less stringent SNR cuts, these planets could be pulled out of \tess data that has already been collected.  The large number of planets in this category could motivate some sort of future search, especially as around 10 of these planets would be around late-type M dwarfs.

We find that 26 -- 37 planets have hosts that have not been observed by \tess yet.  As the extended \tess mission has plans to survey 88\% of the sky overall \citep{Guerrero21}, we can expect that some meaningful percentage of these planets may be discovered in the future when their hosts get surveyed, especially if their periods are short and their hosts are relatively bright.

Finally, we find the existence of 9 -- 13 planets that have not been detected as of the end of Sector 43 (either because of the low SNR or the small number of observed transits), but should have enough observed transits and high enough transit SNRs to be found by the end of Sector 55.  This is the number of planets we expect to find without any extra effort.  While this would represent a significant boost to the overall number of objects in our sample (which currently encompasses 41 short-period rocky planets and 4 rocky planets with $P\,>\,27/2$ days), we would need to observe even more planets to make a precise determination of the planet occurrence rate with our methods.

In general, we note that models with a higher value of $\sigma_A$ corresponds to higher overall planetary occurrence rates and more noise in the individual light curves, resulting in an overall increase in the number of "missed'' planets.  Thus, the $\sigma_A\,=\,0$ yields in Table~\ref{tab:missing_planets} likely represent lower limits.

\section{Summary and Conclusions}
\label{sec:conclusions}
We performed a suite of simulations to explore the observed shape of the rocky M dwarf planet distribution with stellar radius, identifying two key factors that could result in an anomalous distribution shape that were not considered in the previous \textit{TESS}-yield simulations of \cite{Barclay18}.  The first factor is stellar activity.  Estimates of \tess noise depend on the brightness of the star, which is a function of both stellar radius and distance. However, the activity characteristics of small stars also seem to vary with their mass \citep{West15, Robertson13, Astudillo-Defru17}, which may also contribute to the photometric noise.  The second factor may be a non-uniform planet occurrence.  Some studies \citep{Hardegree_Ullman_2019, Mulders_2021} have shown observational and theoretical backing for a mass- or temperature-dependent planet occurrence rate, which could have a complex effect on our final results.

Overall, our simulations, driven by the planetary parameter distributions from \cite{Dressing2015} and methodology from \cite{Barclay18}, mildly overpredict the detection rate of rocky planets around M dwarfs, with more severe overpredictions around the smallest stars ($R_\star\, <\, 0.3 \, R_\odot$).  We also note that the peak of the planet distribution with stellar radius is sharper and occurs at a higher host star radius ($\approx \,0.35\, R_\odot$) than our simulations suggest ($\approx \,0.25 \,R_\odot$).  However, the small number of detected planets in the sample make it difficult to claim anything statistically conclusive about this discrepancy.  

Using \texttt{lightkurve}, we found tentative evidence for a host star radius-dependent term in the CDPP noise of a \tess light curve.  However, we were unable to find any physically-motivated noise model that reproduces the distribution of planets accurately.  We did find that a planet occurrence rate peaking around $0.4\,R_\odot$ and decreasing at lower radii reproduces the observed planet distribution, indicating that we cannot merely extrapolate the previously-observed power-law increase in planet occurrence rate with decreasing radius around larger stars.  

A more extensive study, in line with some of the work done with \kepler \citep[e.g.,][]{Dressing2015}, will be necessary to disentangle precisely which effects are responsible for the observed planet distribution.  Efforts to characterize the sensitivity and completeness of the planet detection pipeline, the false positive rate, and stellar noise would assist in such an analysis. 

As our simulation uncertainties are currently dominated by $\sqrt{N}$ noise, our abilities to draw statistical conclusions about, e.g., the radius-dependent noise or activity of the \tess planets will be improved by surveying more of the sky and the detection of more rocky M dwarf planets.  However, we only expect to detect (optimistically) 13\,$\pm$\,4  additional M dwarf planets through Sector 55 of the \tess mission.  Thus, we may not be able to use the nominal \tess results to precisely constrain the occurrence rate of M dwarfs around late M dwarfs. It might be valuable to re-examine or refine the analysis of the light curves that have already been obtained to try to squeeze out more planet detections. Ultimately, finding all the transiting planets around the latest M dwarfs will probably at least require revising \textit{TESS's} observing strategy, and may even require designing and performing a new survey \citep[e.g.][]{tierras20}.

\vskip 5.8mm plus 1mm minus 1mm
\vskip1sp
We thank David Latham, Rafael Luque, and the anonymous referee for helpful suggestions.

This material is based upon work supported by the National Science Foundation Graduate Research Fellowship under Grant No. DGE 1746045. JLB acknowledges support for this work from the NSF (award number 2108465) and NASA (through the \tess Cycle 4 GI program, grant number 80NSSC22K0117).

This research made use of \texttt{lightkurve}, a Python package for \kepler and \tess data analysis \citep{lightkurve}.  In addition, it has  also used some \texttt{lightkurve} functions with \texttt{astroquery} \citep{astroquery} dependencies.  This research made use of \href{http://www.astropy.org}{\texttt{Astropy}}, a community-developed core Python package for Astronomy \citep{Astropy1, Astropy2}, as well as \texttt{Numpy} \citep{numpy} and \texttt{Scipy} \citep{2020SciPy-NMeth}. This research has made use of NASA's Astrophysics Data System Bibliographic Services.

Some of the data presented in this paper were obtained from the Mikulski Archive for Space Telescopes (MAST) at the Space Telescope Science Institute. The specific observations analyzed can be accessed via \dataset[10.17909/t9-nmc8-f686]{https://doi.org/10.17909/t9-nmc8-f686}.

%\vspace{5mm}
%\nocite{apsrev41Control}
\bibliography{manuscript}
\end{document}